\documentclass[12pt]{article}
\usepackage{amsmath,amssymb,epsfig}

\usepackage{color}
\input{colordvi.tex}

\renewcommand{\thefootnote}{\fnsymbol{footnote}}

\makeatletter

\renewcommand\section{\@startsection {section}{1}{\z@}%
                                   {-3.5ex \@plus -1ex \@minus -.2ex}%
                                   {2.3ex \@plus.2ex}%
                                   {\normalfont\large\bfseries}}

\renewcommand\subsection{\@startsection{subsection}{2}{\z@}%
                                     {-3.25ex\@plus -1ex \@minus -.2ex}%
                                     {1.5ex \@plus .2ex}%
                                     {\normalfont\normalsize\bfseries}}

\newcount\hour \newcount\minute
\hour=\time \divide \hour by 60
\minute=\time
\def\now{%
\ifnum \hour<13
  \ifnum \hour=0 \advance \hour by 12 \number\hour:\else \number\hour:\fi%
     \ifnum \minute<10 0\fi%
     \number\minute%
\ A.M.%
\else \advance \hour by -12 \number\hour:%
  \ifnum \minute<10 0\fi%
  \number\minute%
  \ P.M.%
\fi%
}

\makeatother

\begin{document}

\baselineskip=18pt  
\numberwithin{equation}{section}  
\allowdisplaybreaks  



%
%


\thispagestyle{empty}

\vspace*{-2cm}
\begin{flushright}
YITP-11-08\\
\end{flushright}

\begin{center}

\vspace*{3.0cm}
{\bf\large Higgs Portal to Visible Supersymmetry Breaking}
\vskip 2.0cm
Izawa K.-I.$^{1,2}$,
~Yuichiro Nakai$^{1}$,
~and 
~Takashi Shimomura$^{1}$
\vskip 0.5cm

{\it $^{1}$Yukawa Institute for Theoretical Physics, Kyoto University,\\
     Kyoto 606-8502, Japan}

{\it $^{2}$Institute for the Physics and Mathematics of the Universe, University of Tokyo,\\
     Chiba 277-8568, Japan} \\

\end{center}

\vspace{1.5cm} \centerline{\bf Abstract} \vspace*{0.5cm}

We propose a supersymmetric extension of the standard model whose Higgs
sector induces a spontaneous supersymmetry breaking by itself.
Unlike the minimal extension, the current Higgs mass bound can be evaded
even at the tree-level without the help of
the soft breaking terms due to the usual hidden sector,
as is reminiscent of the next to minimal case.
We also have a possibly light pseudo-goldstino in our visible sector
in addition to extra Higgs particles, both of which stem from supersymmetry breaking dynamics. 
In such a setup of visible supersymmetry breaking,
we may see a part of supersymmetry breaking dynamics
rather directly in future experiments.

\newpage
\setcounter{footnote}{0}
\renewcommand{\thefootnote}{\arabic{footnote}}





\section{Introduction}

The electroweak (EW) symmetry, an $SU(2) \times U(1)$ gauge symmetry,
plays a major role in the standard model of particle physics. In the
standard model, the gauge symmetry is spontaneously broken by the vacuum
expectation value of a Higgs scalar field.
Although the quark-lepton and gauge sectors are well established,
the structure of the Higgs sector is largely uncertain
because the Higgs particle has not yet been discovered directly.
In addition, there is a naturalness problem about the Higgs scalar mass
such that the mass of a scalar field receives large quantum corrections
unlike fermion masses.
Supersymmetry (SUSY) can provide a possible solution to this problem
by introducing the corresponding superpartners into the model with soft SUSY breaking
\cite{Martin:1997ns}.
The minimal supersymmetric extension of the standard model (MSSM) has two Higgs doublets 
\cite{Ibe}
in order to accommodate the anomaly
cancellation and the holomorphicity of the superpotential.

The soft SUSY breaking serves to make
the introduced superpartners heavy enough so that they have not been
observed experimentally. To obtain appropriate soft breaking terms,
the SUSY breaking dynamics is usually put in the so-called hidden sector
that is somehow separated from the visible standard model sector.
Namely, the original SUSY breaking in the hidden sector is mediated
to the visible sector by (flavor-blind) interactions such as gravity or the
standard model gauge interactions.
In the MSSM, the EW symmetry breaking is tied to the resultant SUSY
breaking in the visible sector. It is possibly generated radiatively
through the SUSY breaking mediated from the hidden sector.
If the hidden sector SUSY breaking occurs dynamically
with its breaking scale given by dimensional transmutation,
then the hierarchy between the Planck/GUT scale and the EW scale may be naturally explained.

Unfortunately, this simple scenario is spoiled by the need for
the supersymmetric Higgs mass term called $\mu$-term
\cite{Nel}.
The supersymmetric mass scale must be tuned to about the same size as the EW scale for
the correct symmetry breaking.%
\footnote{
One approach to this problem is to add a singlet superfield whose scalar
component leads to the effective $\mu$-term, which amounts to the
Next to Minimal Supersymmetric Standard Model (NMSSM)
\cite{NMSSM}. Note that when we combine the NMSSM with SUSY-breaking
mediation such as gauge mediation, it is not so easy to
obtain the correct EW symmetry breaking 
\cite{deGouvea:1997cx}.}
Moreover, even if we assume an appropriate order of magnitude for the supersymmetric scale, 
in the MSSM, additional fine-tuning of a few percent is required as follows.
The lightest CP-even Higgs mass $m_h$ is smaller than the $Z$
boson mass at the tree-level in the MSSM. Thus
the current experimental bound $m_h > 114$ GeV
requires large radiative corrections from the (s)top loops
\cite{Okada:1990gg}
with the stop mass of at least $1$ TeV, which in turn affects radiatively
on the soft scalar mass of the up-type Higgs field through the Yukawa coupling.
The soft mass implied by the renormalization is comparable to the stop
mass with a negative sign.
Then, fine-tuning is needed up to a few percent between the $\mu$-term and
the soft scalar mass
of the Higgs field in order to obtain the correct $Z$ boson mass.
Although many solutions to the above problems have been proposed so far,%
\footnote{For example, see \cite{Agashe:1997kn} for relieving the
tension between generation of the $\mu$-term and gauge mediation. See
also \cite{lh} for solving the little hierarchy problem from the view
point of General Gauge Mediation \cite{Meade:2008wd}.}
we do not have any compelling reasons to stick to the minimal (or
next to minimal) Higgs sector like the (N)MSSM and the radiative EW
symmetry breaking driven by the soft SUSY breaking terms.

On the contrary, in this paper,
we regard the
Higgs sector as a window
\cite{Patt}
to unknown physics beyond the MSSM,
in particular, SUSY breaking dynamics.
Historically, visible sector SUSY breaking
\cite{Kumar}
was abandoned due to phenomenological difficulties such as the prediction of light
superpartners, and in turn, hidden sector SUSY breaking has been adopted.
However, in the presence of the hidden sector, additional visible SUSY
breaking is not forbidden phenomenologically.
Namely, we may consider that SUSY breaking is ubiquitous not only
in the hidden sector
\cite{INY}
but also in the visible sector. 

By visible SUSY breaking, we mean the existence of SUSY breaking
in the standard model sector even in the absence of the soft breaking
terms stemming from the usual hidden sector. The SUSY breaking scale of the
hidden sector tends to be too high to observe its dynamics directly in the
foreseeable future. In contrast, if visible SUSY breaking
exists, we may see a part of SUSY breaking dynamics rather directly in
near future experiments.%
\footnote{
Even multiple kinds of extended SUSY breaking might be observable rather directly.
The presence of extra superpartners such as multiple kinds of gravitinos
could open up such a possibility
\cite{Izawa:2010bc}.\label{fn:extended}}
Concretely, as advocated above, we seek visible SUSY breaking in the
Higgs sector, which has large uncertainty at present.
The simplest possibility may be a model that has a singlet field $S$
like the NMSSM with its superpotential coupling to Higgs fields given by
$S H_u H_d$,
where $H_u$ and $H_d$ are the up-type and down-type Higgs superfields.
Then, the vacuum expectation values of the scalar component and the
$F$-term of a visible SUSY breaking field $S$ lead to the effective
$\mu$-term and $B \mu$-term, respectively. These vacuum values are possibly
generated spontaneously by some low-scale dynamics different
from that of the usual SUSY breaking hidden sector. Such a low-scale dynamics
is hopefully within the reach of direct experiments.

It is interesting that we are able to consider even more direct SUSY breaking
dynamics in the visible sector:
the up-type and down-type Higgs fields can participate in the
dynamics of visible SUSY breaking as well as EW symmetry breaking.
That is, if we turn off the standard model gauge interactions and the
soft breaking terms, our Higgs sector reduces to just an O'Raifeartaigh
model with global $SU(2) \times U(1)$ symmetry breaking.
We concentrate on this possibility
below as a concrete example of visible SUSY breaking,
since this model seems advantageous from a perspective of direct
experimental detection.

The rest of the paper goes as follows.
In section $2$, we present our model and analyze its vacuum structure.
Then, in section $3$, we show the mass spectrum of the Higgs sector
in the visible SUSY and EW symmetry breaking vacuum.
It turns out that the lightest CP-even Higgs mass can evade
its current bound even at the tree-level, as is reminiscent of the next
to minimal case.
In section $4$, we discuss a possible connection between the mass
parameters in our Higgs sector and the mass scales of the hidden sector.
Finally, in section $5$, we conclude our discussion and provide possible
directions for future works.

\section{Visible SUSY \& EW breaking}

\begin{table}[tdp]
\begin{center}
\begin{tabular}{c|cccc}
 & $SU(2)_L$ & $U(1)_Y$ & $U(1)_R$
 \\
 \hline
$X_0$ & $\mathbf 1$ & $0$
& $2$
\\
  $X_1$  &  $\mathbf  2$ & $-1/2$
  & $2$
  \\
  $X_2$ & $\mathbf 2$ & $1/2$ & $2$ \\
  $H_u$ & $\mathbf 2$ & $1/2$ & $0$ \\
  $H_d$ & $\mathbf 2$ & $-1/2$ & $0$
\end{tabular}
\end{center}
\caption{The charge assignments of the Higgs sector fields under the EW symmetry and $U(1)_{R}$ symmetry.}
\label{tab:higgs}
\end{table}

Let us first present our model of visible SUSY breaking to provide the scalar potential.
Then, we identify our vacuum in which both of the visible SUSY and the EW symmetry are spontaneously 
broken before analyzing the mass spectrum of the Higgs sector in the vacuum in the next section.

\subsection{The model}
As mentioned in the Introduction, we consider an O'Raifeartaigh model as a mechanism of visible SUSY breaking,
in which an $F$-term of a superfield is non-vanishing.
The minimal extension for this purpose is to introduce a gauge singlet $X_0$ under $SU(2)_L \times U(1)_Y$, 
and a vector-like pair $X_1$, $X_2$ of $SU(2)_L$ doublets%
\footnote{We can also consider an O'Raifeartaigh model with a
vector-like pair of $SU(2)_L$ triplets instead of the doublets, which we
regard as the next to minimal extension and only study the minimal case in
this paper.}
in addition to the usual
up-type and down-type Higgs fields $H_{u,d}$ of the MSSM.%
\footnote{The doubling of the Higgs doublets might be a manifestation
of hidden partial extended SUSY (see also footnote
\ref{fn:extended}). We note that one of the advantages in the minimal (or next to minimal)
Higgs sector like the (N)MSSM may be the gauge coupling unification.
See \cite{Amaldi:1991cn} for discussions on the gauge coupling
unification in the case with four Higgs doublets like the present setup.}
For simplicity, we assume that the model has $U(1)_{R}$ symmetry except for Majorana gaugino masses.%
\footnote{R-symmetric supersymmetric standard model was studied in
\cite{Kribs:2007ac}, whose authors assume that the gauge sector also respects
R-symmetry, so that Majorana gaugino masses are forbidden. In order to give
non-zero masses for the gauginos, they introduce new fields of adjoint
representations under the standard model gauge symmetries, and form the
Dirac gaugino mass terms. Here, just for simplicity of the presentation,
we assume that the gauge sector does not respect R-symmetry, and hence
Majorana gaugino mass terms are allowed.
It is straightforward to extend our model to include the Dirac mass
terms to preserve $U(1)_{R}$ symmetry by introducing additional fields
of the adjoint representations under the standard model gauge
group. Then, the supersymmetric flavor
problems may be ameliorated, as pointed out in \cite{Kribs:2007ac}.
}
The charge assignments of the Higgs sector fields
under the EW symmetry and $U(1)_{R}$ symmetry are summarized in
Table~\ref{tab:higgs}.
We assign R-charge $1$ for all the matter superfields,%
\footnote{This assignment allows Majorana neutrino mass terms $H_u L H_u L$.}
so that it forbids renormalizable superpotential terms such as
$Q L \bar{d} + L L \bar{e} + L H_u + \bar{d} \bar{d} \bar{u}$,
which violate the lepton or baryon number.
Apart from the usual Yukawa couplings of Higgs fields $H_{u,d}$ with matters,
the symmetries allow our superpotential to have the following terms:
\begin{equation}
W_{Higgs}= X_0 \left(f+\lambda H_u H_d \right) + m_1 X_1 H_u + m_2 X_2 H_d,
\label{visibleW}
\end{equation}
where a coupling $f$ has mass dimension $2$, and $m_1, m_2$ have mass dimension $1$.
We can take all these couplings real without loss of generality.
All the mass scales are assumed to be of order the EW scale.

With the canonical K\"ahler potential of all the fields,
the superpotential and the gauge interactions
determine the scalar potential of the Higgs sector. 
The entire scalar potential of the Higgs sector consists of $F$-terms, $D$-terms and the soft SUSY breaking terms:
\begin{equation}
V_{Higgs} = V_F + V_D + V_{soft}.
\label{visibleV}
\end{equation}
{}From \eqref{visibleW}, the $F$-term contribution to the scalar potential is given by
\begin{equation}
\begin{split}
V_F &=  \left| f+ \lambda H_u^+ H_d^- - \lambda H_u^0 H_d^0 \right|^2 \\
&\quad + m_1^2 \left( | H_u^0 |^2 + | H_u^+ |^2 \right) + m_2^2 \left( | H_d^0 |^2 + | H_d^- |^2 \right) \\
&\quad + \left| \lambda X_0 H_d^0 - m_1 X_1^0 \right|^2 + \left| \lambda X_0 H_d^- - m_1 X_1^- \right|^2 \\
&\quad + \left| \lambda X_0 H_u^0 + m_2 X_2^0 \right|^2 + \left| \lambda X_0 H_u^+ + m_2 X_2^+ \right|^2, \\
\end{split}\label{vf}
\end{equation}
where the superscripts of the fields denote the electric charges.
On the other hand, from Table~\ref{tab:higgs},
we can derive the following $D$-term contribution of the Higgs sector:
\begin{equation}
V_D = \frac{1}{2} D_2^a D_2^a + \frac{1}{2} D_1D_1,
\end{equation}
where $D_2^a~(a=1, 2, 3)$ and $D_1$ represent the contributions of the
Higgs sector to the $D$-terms of $SU(2)_L$ and $U(1)_Y$ vector
superfields, and the summation over $a$ should be understood. $D$-terms
involving only the Higgs
fields are given by 
\begin{equation}
\begin{split}
D_2^a &= -g_2 \left( H_u^{\ast} \tau^a H_u + H_d^{\ast} \tau^a H_d + X_1^{\ast} \tau^a X_1 + X_2^{\ast} \tau^a X_2 \right), \\
D_1 &= -\frac{g_1}{2} \left(| H_u^0 |^2 + | H_u^+ |^2 - | H_d^0 |^2 - | H_d^- |^2 - | X_1^0 |^2 - | X_1^- |^2 + | X_2^0 |^2 + | X_2^+ |^2\right).
\end{split}
\end{equation}
where $g_2$ and $g_1$ are the gauge couplings of $SU(2)_L$ and $U(1)_Y$, and $\tau^a$ denote $SU(2)_L$ generators.
The soft SUSY breaking terms for the Higgs
fields are the usual ones mediated from the hidden sector.
The soft terms which respect the symmetries are given as follows:%
\footnote{In particular, the $U(1)_{R}$ symmetry makes $A$-terms vanishing.}
\begin{equation}
\begin{split}
V_{soft} &= m_{H_u}^2(| H_u^0 |^2 + | H_u^+ |^2) + m_{H_d}^2(| H_d^0 |^2 + | H_d^- |^2) + m_{X_0}^2 |X_0|^2 \\
&\quad + m_{X_1}^2 (| X_1^0 |^2 + | X_1^- |^2) + m_{X_2}^2 (| X_2^0 |^2 + | X_2^+ |^2) \\
&\quad + b (H_u^+ H_d^- - H_u^0 H_d^0) + c.c.,
\end{split}\label{vsoft}
\end{equation}
where $m^2_i~(i=H_u,H_d,X_0,X_1,X_2)$ are soft scalar masses of the
fields and $b$ is a bilinear coupling for the Higgs fields.

\subsection{Our vacuum}

We now specify our vacuum to minimize the above potential.
In order to demonstrate the idea of visible SUSY breaking (in the Higgs sector),
we first analyze the limit of turning off the standard model gauge
interactions and the soft breaking terms \eqref{vsoft}.
Then, the model \eqref{visibleW} just reduces to an O'Raifeartaigh model
with global $SU(2) \times U(1)$ symmetry spontaneously broken,\footnote{We temporarily require a coupling relation $\lambda f > m_1 m_2$ in this limit. In contrast, this kind of O'Raifeartaigh models as a hidden sector \cite{gauge} requires $\lambda f < m_1 m_2$ in order to obtain a SUSY breaking vacuum without the gauge symmetry breaking.}
so that it is enough to deal with the $F$-term contribution \eqref{vf}.
We assume that the vacuum expectation values of all the electrically charged
fields are vanishing, which will be justified retrospectively by the mass spectrum around the
vacuum. Then, the scalar potential is written as
\begin{equation}
\begin{split}
V_F &=  \left| f - \lambda H_u^0 H_d^0 \right|^2 + m_1^2 | H_u^0 |^2 + m_2^2 | H_d^0 |^2 \\
&\quad + \left| \lambda X_0 H_d^0 - m_1 X_1^0 \right|^2 + \left| \lambda X_0 H_u^0 + m_2 X_2^0 \right|^2.
\end{split}\label{vfc0}
\end{equation}
We emphasize here that the $F$-terms of all the neutral fields
cannot be simultaneously taken to be zero in the vacuum, and hence SUSY is
spontaneously broken in the Higgs sector. Since the soft SUSY breaking
terms have been turned off, SUSY is broken in the visible 
sector by itself. This is, what we call, the visible SUSY breaking in
the present scenario.

First, let us consider the minimization of the above scalar potential
with respect to $X_0$,
\begin{equation}
\frac{\partial V}{\partial X_0^\ast} = \lambda {H_d^0}^\ast \left( \lambda X_0 H_d^0 - m_1 X_1^0 \right) +  \lambda {H_u^0}^\ast \left( \lambda X_0 H_u^0 + m_2 X_2^0 \right) = 0.
\label{miniX0l}
\end{equation}
We can choose $X_0 = X_1^0 = X_2^0 =0$ as a solution to this equation,
which also satisfies the similar minimization conditions about $X_1^0$ and $X_2^0$.
Next, we proceed to the minimization about the ordinary Higgs fields $H_u^0,
H_d^0$. The vacuum conditions are given by
\begin{equation}
\begin{split}
\frac{\partial V}{\partial {H_u^{0}}^\ast} &=  - \lambda {H_d^0}^\ast \left( f - \lambda H_u^0 H_d^0 \right) + m_1^2 H_u^0 = 0, \\
\frac{\partial V}{\partial {H_d^{0}}^\ast} &=  - \lambda {H_u^0}^\ast \left( f - \lambda H_u^0 H_d^0 \right) + m_2^2 H_d^0 = 0, \\
\end{split}
\end{equation}
where we have used the solution $X_0 = X_1^0 = X_2^0 =0$.
Up to symmetry rotation,
the expectation values of the fields $H_u^0, H_d^0$ can be taken to be
real. Then, the above conditions can be solved as follows:
\begin{equation}
H_u^{0} =  \frac{1}{\lambda} \sqrt{\frac{m_2}{m_1} \left( \lambda f - m_1 m_2 \right)}, \qquad H_d^{0} =  \frac{1}{\lambda} \sqrt{\frac{m_1}{m_2} \left( \lambda f - m_1 m_2 \right)},
\end{equation}
where the global $SU(2) \times U(1)$ symmetry is broken to the remaining
$U(1)$ symmetry. When the global symmetry is gauged as is done in the standard
model, this corresponds to the EW symmetry breaking.

We are now in a position to analyze the full scalar potential
\eqref{visibleV} and specify our vacuum in which SUSY and the EW symmetry are broken. 
As described above,
we here assume that the vacuum values of all the electrically charged
fields are vanishing. Then, the relevant scalar potential is given by
\begin{equation}
\begin{split}
V &=  \left| f - \lambda H_u^0 H_d^0 \right|^2 + \left| \lambda X_0 H_d^0 - m_1 X_1^0 \right|^2 + \left| \lambda X_0 H_u^0 + m_2 X_2^0 \right|^2 \\
&\quad + \mu_1^2 | H_u^0 |^2  + \mu_2^2 | H_d^0 |^2 - \left( b H_u^0 H_d^0 + c.c. \right) \\
&\quad + m_{X_0}^2 |X_0|^2 + m_{X_1}^2 | X_1^0 |^2 + m_{X_2}^2 | X_2^0 |^2 \\
&\quad +\frac{1}{8} g^2 \left(| H_u^0 |^2 - | H_d^0 |^2 - | X_1^0 |^2 + | X_2^0 |^2 \right)^2,
\end{split}
\end{equation}
where we have defined mass parameters $\mu_1^2 = m_1^2 + m_{H_u}^2$,
$\mu_2^2 = m_2^2 + m_{H_d}^2$, and a coupling $g^2 = g_1^2 + g_2^2$ to
simplify the expression. Although the minimization condition about $X_0$ is
slightly changed from \eqref{miniX0l} by the soft scalar mass term of
$X_0$, we can keep choosing $X_0 = X_1^0 = X_2^0 =0$ as a solution
which simultaneously satisfies the minimization conditions about $X_1^0$ and $X_2^0$.
Next, we consider the minimization conditions about the Higgs fields $H_u^0$ and $H_d^0$.
We can again take the expectation values of these fields real without loss of generality, and express them as 
$H_u^0 = \frac{1}{\sqrt{2}} v \sin \beta$ and $H_d^0 =
\frac{1}{\sqrt{2}} v \cos \beta$, as is done in the case of the MSSM.
These vacuum values break the EW gauge symmetry
to produce masses for the $W$ bosons and the $Z$ boson,
\begin{equation}
m_W^2 = \frac{1}{4} g_2^2 \, v^2, \quad m_Z^2 = \frac{1}{4} g^2 v^2,
\end{equation}
where $v^2 \simeq (246 \, \mathrm{GeV})^2$ is required 
in order to obtain the observed values of the masses. Then, the minimization conditions
$\partial V / \partial H_u^{0} = \partial V / \partial H_d^{0} = 0$
result in the following expressions:
\begin{equation}
\begin{split}
&\mu_1^2 + \frac{1}{2} \lambda^2 v^2 \cos^2 \beta  = \left(\lambda f + b \right) \cot \beta + \frac{m_Z^2}{2} \cos 2\beta, \\
&\mu_2^2 + \frac{1}{2} \lambda^2 v^2 \sin^2 \beta  = \left(\lambda f + b \right) \tan \beta - \frac{m_Z^2}{2} \cos 2\beta.
\end{split}\label{vev}
\end{equation}
Note that these conditions are very similar to the ones in the case of the 
MSSM. In fact, if we take the limit $\lambda \rightarrow 0$, the conditions appear
the same as the corresponding equations of the MSSM. In this limit or in the MSSM, the soft SUSY
breaking terms are essential for the correct EW symmetry breaking \cite{Batra:2008rc}.
On the other hand, in our model, the correct symmetry breaking is realized even in the absence of the soft breaking 
terms for nonzero $\lambda$,
since the effects of the soft SUSY breaking terms
are solely contained in the expressions through the forms
$\mu_1^2 = m_1^2 + m_{H_u}^2$, $\mu_2^2 = m_2^2 + m_{H_d}^2$ and $\lambda f + b$.

By means of \eqref{vev}, we obtain the following
expression of the $Z$ boson mass in terms of the mass parameters $\mu_1$ and $\mu_2$:
\begin{equation}
m_Z^2 = - \left( \frac{\mu_2^2 - \mu_1^2}{\cos 2 \beta} + \mu_2^2 + \mu_1^2 \right).
\label{veq}
\end{equation}
As will be shown in the next section, in this model, we can obtain the lightest CP-even Higgs mass $m_h$ so as
to evade the current mass bound $m_h > 114$ GeV%
\footnote{We simply adopt this value for the Higgs boson
in the standard model as a point of reference also in our estimate,
though it does not necessarily apply in our case.}
even at the tree-level, as is reminiscent of the NMSSM.
Thus, we do not need large soft scalar masses beyond $1$ TeV
to get large radiative corrections. Namely, the mass parameters $\mu_1,\mu_2$
can be near the EW scale, so that lesser fine-tuning is required
to obtain the correct $Z$ boson mass in the above equation.

\section{Mass spectrum}

In this section, we show the mass spectrum of the Higgs sector
fields in the visible SUSY and EW symmetry breaking vacuum discussed
above. We first analyze the scalar masses. It turns out that
the lightest CP-even Higgs mass can be above the current mass bound even
at the tree-level, as is reminiscent of the NMSSM case.
Then, we move to the discussion of the fermion masses.
One of the neutralinos is massless at the tree-level,
which would correspond to the goldstino in the visible SUSY breaking
without soft SUSY breaking terms. 

\subsection{The scalar masses}

The scalar fields of the Higgs sector consist of $18$ real field degrees
of freedom. When the EW symmetry is broken, three of them are the
would-be Nambu-Goldstone bosons which are eaten by the $Z$ and the $W^\pm$.
The remaining $15$ of them are the physical modes. We now expand the Higgs fields 
around their vacuum expectation values as
\begin{equation}
\begin{split}
H_u^0 &\to \frac{1}{\sqrt{2}} v \sin \beta + H_u^0, \\
H_d^0 &\to \frac{1}{\sqrt{2}} v \cos \beta + H_d^0,
\end{split}\label{shift}
\end{equation}
where the dynamical parts are further decomposed into CP-even and odd ones as follows:
\begin{equation}
H_u^0 = \frac{1}{\sqrt{2}} \left( \eta_1 + i \xi_1 \right), \qquad H_d^0 = \frac{1}{\sqrt{2}} \left( \eta_2 + i \xi_2 \right).
\end{equation}
Here, $\eta_{1,2}$ are CP-even scalar fields and $\xi_{1,2}$ are CP-odd ones. 

First, we analyze the masses of the CP-odd parts. From \eqref{visibleV},
we can read the mass terms of the corresponding fields.%
\footnote{Their expressions are summarized in the Appendix.\label{fn:see-app}}
The mass matrix for $\xi_1$ and $\xi_2$ is given by
\begin{equation}
\begin{split}
{\cal M}_{\xi}^2 = \frac{1}{2}
  \left(
  \begin{array}{cc}
  \xi_1, & \xi_2
  \end{array}
  \right)
\left(
  \begin{array}{cc}
  \left( \lambda f + b \right) \cot \beta & \lambda f + b \\
  \lambda f + b & \left( \lambda f + b \right) \tan \beta
  \end{array}
 \right)
 \left(
  \begin{array}{c}
  \xi_1 \\
\xi_2
  \end{array}
 \right),
\end{split}
\end{equation}
which takes the same form as that of the MSSM
except for the $\lambda f$ terms.
Diagonalizing this matrix, the eigenvalues turn out to be
\begin{equation}
m_{\chi^0}^2 = 0, \qquad m_{A^0}^2 = \mu_1^2 + \mu_2^2 + \frac{1}{2} \lambda^2 v^2 = \frac{2 \left( \lambda f + b\right)}{\sin 2\beta}.
\label{Amass}
\end{equation}
The massless field is the would-be Nambu-Goldstone mode eaten by the $Z$
boson. The corresponding mass eigenstates are expressed as
\begin{equation}
\chi^0 = \xi_1 \sin \beta - \xi_2 \cos \beta, \qquad A^0 = \xi_1 \cos \beta + \xi_2 \sin \beta.
\end{equation}

\begin{figure}[t]
\begin{tabular}{cc}
\includegraphics[width=80mm]{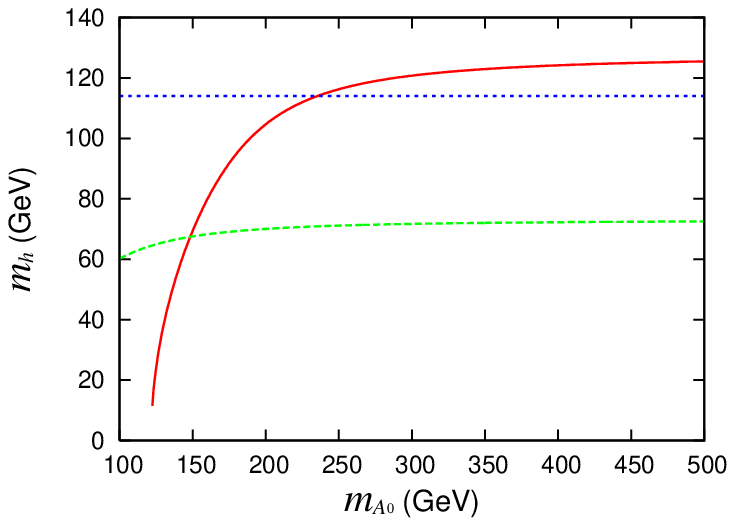}
 &
 \includegraphics[width=80mm]{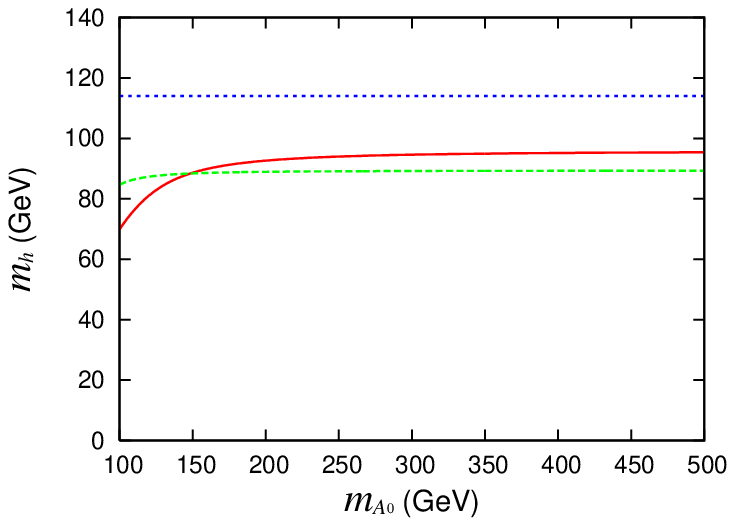}
\end{tabular}
\vspace{-0.3cm}
\caption{The mass of the lighter Higgs, $m_h$, in our model (red (solid) curves). The horizontal axis is the mass of $A^0$. The left panel is plotted with $\tan \beta = 3$, while the right panel is done with $\tan \beta = 10$. The horizontal (dotted) line denotes the current Higgs mass bound. The green (dashed) curve represents the case of the MSSM.}
\label{fig:hmass}
\end{figure}

Next, we investigate the masses of the CP-even parts $\eta_1$ and
$\eta_2$ of the neutral Higgs fields.
The analysis of the mass terms proceeds in the same way as above.${}^{\ref{fn:see-app}}$
The mass matrix is given by
\begin{equation}
\begin{split}
{\cal M}_{\eta}^2 = \frac{1}{2}
\left(
  \begin{array}{cc}
  \eta_1, & \eta_2
  \end{array}
 \right)
\left(
  \begin{array}{cc}
  m_{A^0}^2 \cos^2 \beta + m_Z^2 \sin^2 \beta & \left( \lambda^2 v^2 - m_{A^0}^2 - m_Z^2 \right) \sin \beta \cos \beta \\
  \left( \lambda^2 v^2 - m_{A^0}^2 - m_Z^2 \right) \sin \beta \cos \beta & m_{A^0}^2 \sin^2 \beta + m_Z^2 \cos^2 \beta
  \end{array}
 \right)
 \left(
  \begin{array}{c}
  \eta_1 \\
\eta_2
  \end{array}
 \right).
\end{split}
\label{neutral_higgs_matrix}
\end{equation}
Then, the eigenvalues of this mass matrix are given by
\begin{equation}
m_{h,H}^2 = \frac{1}{2} \left( m_{A^0}^2 + m_Z^2 \mp \sqrt{\left( m_{A^0}^2 - m_Z^2 \right)^2 + 4 \left( m_{A^0}^2 - \frac{1}{2} \lambda^2 v^2 \right) \left( m_{Z}^2 - \frac{1}{2} \lambda^2 v^2 \right) \sin^2 2\beta } \right), \label{neutral_higgs_mass}
\end{equation}
which also take the same forms as those in the MSSM except for the terms
dependent on $\lambda$. Note that this slight difference is,
nonetheless, crucial for the lighter CP-even Higgs mass to evade the
current experimental bound, as is the case for the NMSSM. In fact, 
in the limit of large $m_{A^0}$, the lighter Higgs mass can be written as 
\begin{align}
m_h^2 \simeq m_Z^2 \cos^2 2\beta + \frac{1}{2}\lambda^2 v^2 \sin^2 2\beta, \label{appr_higgs_mass}
\end{align}
which is lifted up by the second term in the right-hand side for large $\lambda$ and small $\tan\beta$.

Let us now analyze the masses of the charged Higgs fields.
The analysis of the mass terms again proceeds in the same way.${}^{\ref{fn:see-app}}$
The mass matrix for the charged Higgs fields is given by
\begin{equation}
\begin{split}
{\cal M}_{H^{\pm}}^2 = \left( \mu_1^2 + \mu_2^2 + M_W^2 \right)
\left(
  \begin{array}{cc}
  {H_u^+}^\ast, & H_d^-
  \end{array}
 \right)
\left(
  \begin{array}{cc}
  \cos^2 \beta & \sin \beta \cos \beta \\
  \sin \beta \cos \beta & \sin^2 \beta
  \end{array}
 \right)
 \left(
  \begin{array}{c}
  H_u^+ \\
{H_d^-}^\ast
  \end{array}
 \right).
\end{split}
\end{equation}
Then, the eigenvalues of this mass matrix are obtained as
\begin{equation}
m_{\chi^\pm}^2 = 0, \qquad m_{H^\pm}^2 = \mu_1^2 + \mu_2^2 + M_W^2 = m_{A^0}^2 + M_W^2 - \frac{1}{2} \lambda^2 v^2, 
\label{charged_higgs_mass}
\end{equation}
where $\chi^- = {\chi^+}^\ast$ and $H^- = {H^+}^\ast$.
The massless modes $\chi^\pm$ are would-be Nambu-Goldstone modes eaten by the $W$ boson.
We also note that the mass relation between the $A^0$ mass and the
masses of the $H^\pm$ coincides with that of the MSSM
except for the term dependent on the coupling $\lambda$. The mass eigenstates are given by
\begin{equation}
\chi^+ = H_u^+ \sin \beta - {H_d^-}^\ast \cos \beta, \qquad H^+ = H_u^+ \cos \beta + {H_d^-}^\ast \sin \beta.
\end{equation}

\begin{figure}[t]
\begin{tabular}{cc}
\includegraphics[width=80mm]{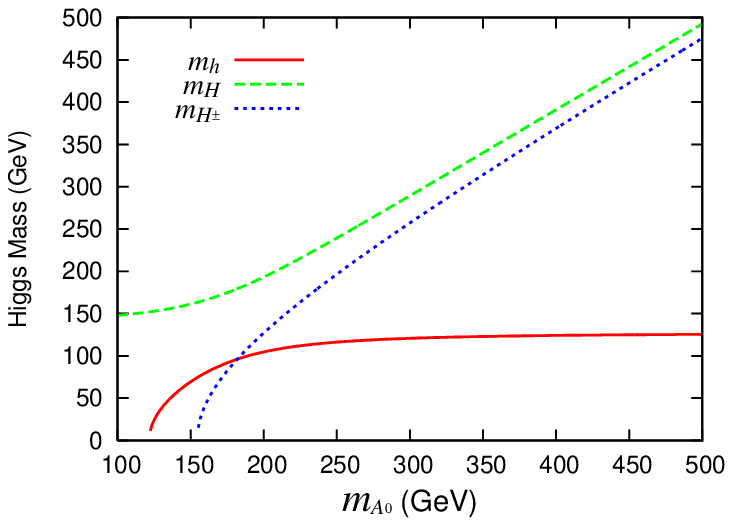}
 &
 \includegraphics[width=80mm]{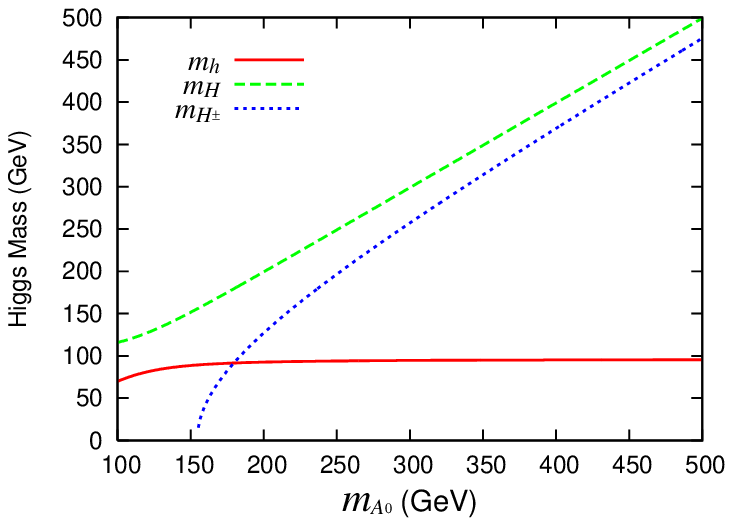}
\end{tabular}
\vspace{-0.3cm}
\caption{The lighter Higgs mass $m_h$ (red (solid) curves), the heavier Higgs mass $m_H$ (green (dashed) curves), and the charged Higgs mass $m_{H^\pm}$ (blue (dotted) curves). The horizontal axis is the mass of $A^0$. The left panel is plotted with $\tan \beta = 3$, while the right panel is done with $\tan \beta = 10$.}
\label{fig:hmasses}
\end{figure}

\begin{figure}[p]
\begin{tabular}{cc}
\includegraphics[width=80mm]{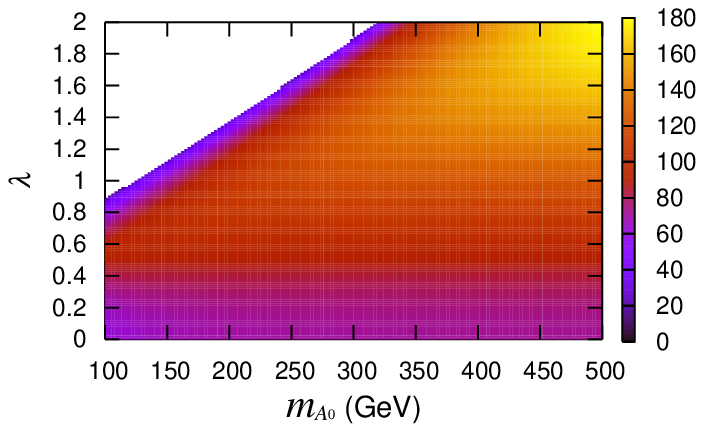}
 &
 \includegraphics[width=80mm]{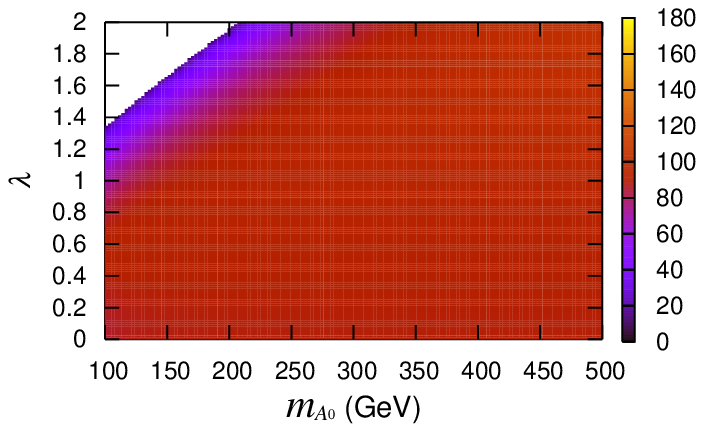}
\end{tabular}
\vspace{-0.5cm}
\caption{The lighter CP-even Higgs mass $m_h$. The horizontal axis is the mass of $A^0$ and the vertical axis is the coupling $\lambda$. The left panel is plotted with $\tan \beta = 3$, while the right panel is done with $\tan \beta = 10$.}
\label{fig:hmasslplot}
\end{figure}

\begin{figure}
\begin{tabular}{cc}
\includegraphics[width=80mm]{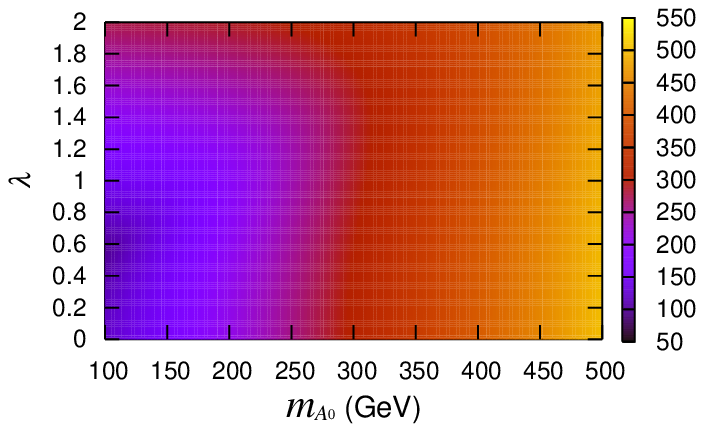}
&
\includegraphics[width=80mm]{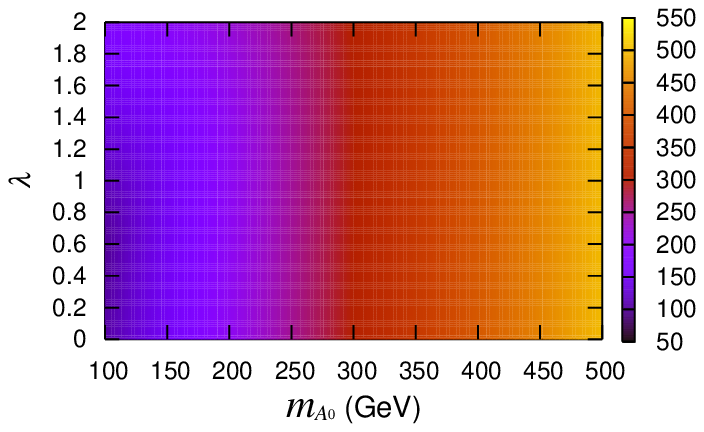}
\end{tabular}
\vspace{-0.5cm}
\caption{The heavier Higgs mass $m_H$. The left panel is plotted with $\tan \beta = 3$, while the right panel is done with $\tan \beta = 10$.}
\label{fig:Hmasslplot}
\end{figure}

\begin{figure}
\begin{tabular}{cc}
\includegraphics[width=80mm]{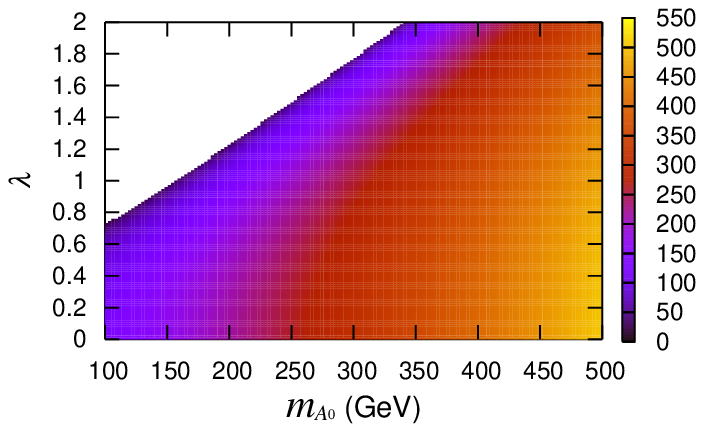}
&
\includegraphics[width=80mm]{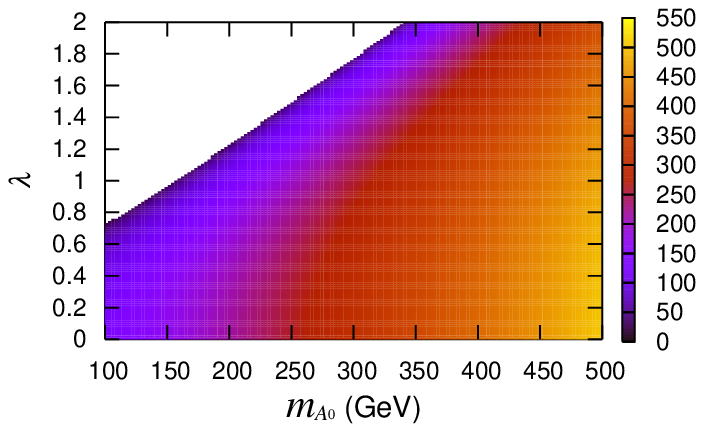}
\end{tabular}
\vspace{-0.5cm}
\caption{The charged Higgs masses $m_{H^\pm}$. The left panel is plotted with $\tan \beta = 3$, while the right panel is done with $\tan \beta = 10$.}
\label{fig:chmasslplot}
\end{figure}

Figure \ref{fig:hmass} shows the mass of the lighter CP-even Higgs, $m_h$, by varying the $A^0$ (red (solid) curves) for $\lambda=1$
in the cases with $\tan \beta = 3$ (the left panel) and with $\tan \beta = 10$ (the right panel). 
The horizontal (dotted) lines represent the current
experimental bound on the Higgs mass, $m_h > 114$ GeV. For comparison, the mass of the lighter Higgs in the MSSM 
is shown (green (dashed) curves), in which $\lambda$ is taken to be zero.
In the left panel, we see that the lighter CP-even Higgs mass can reach above the current
experimental bound for $m_{A^0} > 220$ GeV in our model with $\tan \beta = 3$ unlike the MSSM case.
We also  see in the right panel that the Higgs mass in our model approaches that of the MSSM as $\tan \beta$ is increased.
This behavior can be understood by means of \eqref{appr_higgs_mass}.

In figure \ref{fig:hmasses}, we show the behavior of the masses of the
lighter Higgs, $m_h$ (red (solid) curves), the heavier Higgs, $m_H$
(green (dashed) curves), and the charged Higgs, $m_{H^\pm}$ (blue
(dotted) curves),  in terms of $m_{A^0}$ for $\lambda=1$. Here, $\tan
\beta$ is fixed to $3$ (the left panel) and $10$ (the right panel),
respectively. Both the panels imply that the masses of the charged Higgs are tachyonic for $m_{A^0}$ smaller than $150$ GeV. This is due to the term dependent on $\lambda$ in \eqref{charged_higgs_mass}. 
Similarly, one sees that the mass of the lighter Higgs becomes tachyonic
for $\tan\beta=3$ (the left panel) when $m_{A^0}$ is smaller than $130$
GeV, while it does not for $\tan\beta=10$ (the right panel). This is
because the terms dependent on $\lambda$ in \eqref{neutral_higgs_mass} are proportional to $\sin 2\beta$, which become smaller as $\tan\beta$ becomes larger. Their negative contribution to the lighter Higgs mass is small 
for larger $\tan\beta$, and hence the mass is positive for $m_{A^0} < 130$ GeV with $\tan \beta = 10$.

To see the dependence of the masses on the coupling $\lambda$, we show the contour plots of the masses, $m_h$ (figure \ref{fig:hmasslplot}), $m_H$ (figure \ref{fig:Hmasslplot}), and $m_{H^\pm}$ (figure \ref{fig:chmasslplot}). In each figure, $\tan \beta = 3$ for the left panel, $\tan \beta = 10$ for the right panel, and values of the masses are indicated by a color bar aside. In figure \ref{fig:hmasslplot}, we see that the lighter Higgs mass becomes tachyonic (white region) for large $\lambda$ and small $m_{A^0}$. This is because the second term in the right-hand side of \eqref{neutral_higgs_mass} becomes larger than the first term as $\lambda$ is large.
The region of tachyonic mass for $\tan\beta = 10$ is smaller than that
for $\tan\beta=3$, since the $\lambda$ dependent terms are suppressed by
$\sin^2 2 \beta$. In the left panel of figure \ref{fig:hmasslplot}, it
is seen that the lighter Higgs mass exceeds the current experimental
bound  in a large region of $\lambda > 0.6$.  In figure
\ref{fig:Hmasslplot}, we can see that the heavier Higgs mass is less
sensitive to $\lambda$ and mainly determined by $m_{A^0}$. The terms
dependent on $\lambda$ are significant only in a region of small
$m_{A^0}$ and large $\lambda$. However, such a region is excluded by the
lighter Higgs mass to be smaller than the experimental bound (or even tachyonic).
In figure \ref{fig:chmasslplot}, similarly to the lighter Higgs mass,
one sees that the charged Higgs mass becomes tachyonic for large
$\lambda$ (white region).
To avoid the tachyonic mass, one can obtain an upper bound on $\lambda$
from \eqref{charged_higgs_mass} as
\begin{equation}
 \lambda < g \sqrt{\frac{m_{A^0}^2 + m_W^2}{2 m_Z^2}}.
\end{equation}
This bound is independent of $\tan\beta$ and therefore gives a stronger
constraint on $\lambda$ than that by the lighter Higgs
mass.\footnote{This bound is conservative because we only require that
$m_{H^\pm}$ is positive. When we take into account the current
experimental bound on $m_{H^\pm}$, a more stringent constraint is obtained.}

\begin{figure}[t]
\begin{tabular}{cc}
\includegraphics[width=80mm]{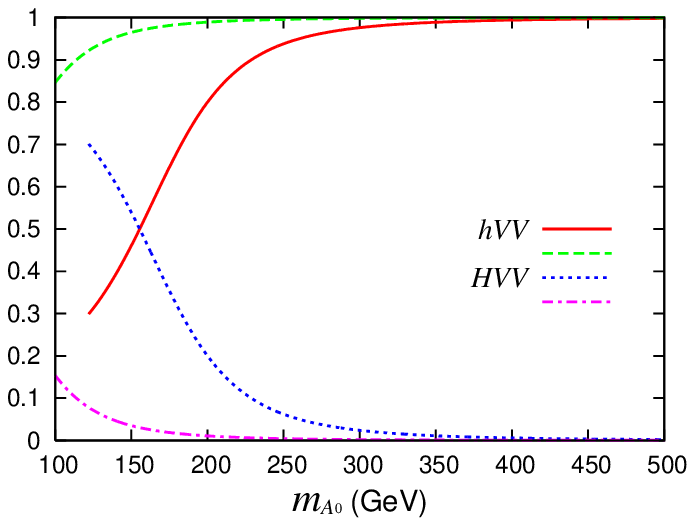}
&
\includegraphics[width=80mm]{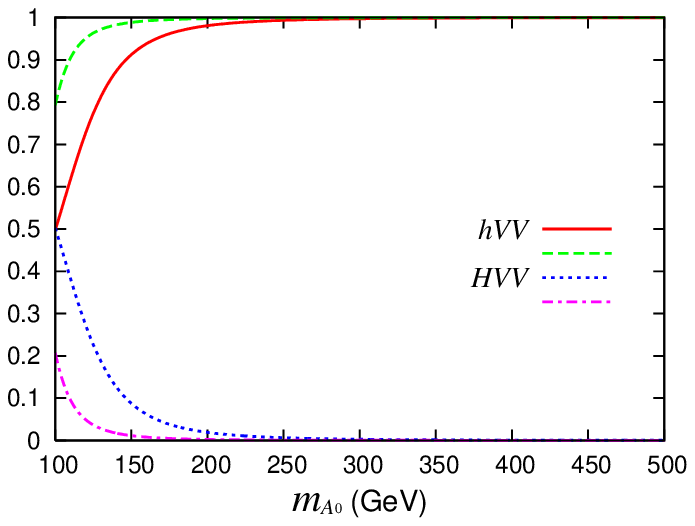}
\end{tabular}
\vspace{-0.3cm}
\caption{The neutral CP-even Higgs boson couplings with the standard model gauge fields. The left graph is plotted with $\tan \beta = 3$, while the right graph is done with $\tan \beta = 10$.}
\label{fig:coupl}
\end{figure}

We now express the mass eigenstates of the neutral CP-even Higgs fields
in terms of the mixing angle $\alpha$ in the same way as is often done
in the analysis of the MSSM as follows:
\begin{equation}
h = \eta_1 \cos \alpha - \eta_2 \sin \alpha, \qquad H = \eta_1 \sin \alpha + \eta_2 \cos \alpha.
\end{equation}
Here, $h$ corresponds to the lighter mass eigenstate, while $H$ corresponds to the heavier one. The mixing angle $\alpha$ is given by
\begin{equation}
\frac{\sin 2\alpha}{\sin 2\beta} = - \frac{m_{A^0}^2 + m_Z^2 - \lambda^2 v^2}{m_H^2 - m_h^2}, \qquad \frac{\cos 2\alpha}{\cos 2\beta} = - \frac{m_{A^0}^2 - m_Z^2}{m_H^2 - m_h^2},
\end{equation}
where the first relation takes the same form as that of the MSSM except
for the $\lambda^2 v^2$ term, and the second one exactly coincides with
that of the MSSM. In order to identify which Higgs boson is the
standard-model-like one, we need to know which Higgs boson is more strongly coupled with the standard model gauge bosons and the matter fields. Higgs--gauge boson--gauge boson couplings are expressed as follows:
\begin{equation}
\begin{split}
&\mathcal{L}_{hgg} = -g_2 m_W \sin (\alpha - \beta) h W_{\mu}^\dagger W^\mu -\frac{1}{2} g m_Z \sin (\alpha - \beta) h Z_\mu Z^\mu \\
&\qquad+ g_2 m_W \cos (\alpha - \beta) H W_{\mu}^\dagger W^\mu +\frac{1}{2} g m_Z \cos (\alpha - \beta) H Z_\mu Z^\mu.
\end{split}
\end{equation}
Figure \ref{fig:coupl} shows the strength of these couplings for
$\lambda = 1$ by varying the mass of the $A^0$ in the case with $\tan \beta = 3$ (the left
panel) and $\tan \beta = 10$ (the right panel). The red (solid) curve
represents the coupling of the lighter Higgs $h$ with the gauge fields
(by $\sin (\alpha - \beta)$),
while the blue (dotted) curve represents that of the heavier Higgs
$H$ (by $\cos (\alpha - \beta)$). For comparison, we also show the same couplings in the case of the MSSM. The green
(dashed) curve represents the coupling of the lighter Higgs in the MSSM,
while the pink (dashed-dotted) curve denotes that of the heavier Higgs
field. From the figure, we see that the coupling of the lighter Higgs in our model is smaller than that of the MSSM for small $m_{A^0}$, and 
can be in anti-decoupling region for $m_{A^0} < 180$ GeV in the $\tan\beta=3$ case. We also see that the lighter CP-even Higgs field $h$
is more strongly coupled with the standard model gauge fields than the heavier one $H$ in the region of large $A^0$ mass, which 
is the same as in the MSSM.

Finally, we analyze the masses of the $X$ scalar fields. Their mass
terms derived from \eqref{visibleV} are also summarized in the Appendix. The masses of the charged fields are given by
\begin{equation}
\begin{split}
m_{X_1^-}^2 &= m_1^2 + m_{X_1}^2 - \frac{1}{2} m_Z^2 \cos 2\theta_W \cos 2\beta, \\
m_{X_2^+}^2 &= m_2^2 + m_{X_2}^2 + \frac{1}{2} m_Z^2 \cos 2\theta_W \cos 2\beta,
\end{split}
\end{equation}
where $\theta_W$ denotes the Weinberg angle, $\sin^2 \theta_W \simeq 0.23$. In order to analyze the masses of the neutral fields, we define
\begin{equation}
\begin{split}
&X_0 = \frac{1}{\sqrt{2}} (\sigma_0 + i \rho_0), \\
&X_1^0 = \frac{1}{\sqrt{2}} (\sigma_1 + i \rho_1), \\
&X_2^0 = \frac{1}{\sqrt{2}} (\sigma_2 + i \rho_2).
\end{split}
\end{equation}
Then, the mass matrix of the real parts $\sigma$ is expressed as
\begin{equation}
\begin{split}
{\cal M}_{\sigma}^2 = \frac{1}{2}
\left(
  \begin{array}{ccc}
  \sigma_0 & \sigma_1 & \sigma_2
  \end{array}
 \right)
\left(
  \begin{array}{ccc}
  \frac{1}{2} \lambda^2 v^2 + m_{X_0}^2 & -\frac{\lambda}{\sqrt{2}} m_1 v c_\beta & \frac{\lambda}{\sqrt{2}} m_2 v s_\beta \\
  -\frac{\lambda}{\sqrt{2}} m_1 v c_\beta & m_1^2 + m_{X_1}^2 + \frac{1}{2} m_Z^2 c_{2\beta} & 0 \\
  \frac{\lambda}{\sqrt{2}} m_2 v s_\beta & 0 &  m_2^2 + m_{X_2}^2 - \frac{1}{2} m_Z^2 c_{2\beta}
  \end{array}
 \right)
 \left(
  \begin{array}{c}
  \sigma_0 \\
\sigma_1 \\
\sigma_2
  \end{array}
 \right),
\end{split}\label{sigmass}
\end{equation}
where $s_\beta$, $c_\beta$ and $c_{2\beta}$ denote $\sin \beta$, $\cos
\beta$ and $\cos2\beta$, respectively. As for the imaginary
parts $\rho$, their mass matrix takes the same form as the real part
mass matrix \eqref{sigmass}. We present  sample spectra of the $X$
scalar masses in table~\ref{tab1}, where the following four cases with
$\tan \beta = 3$ (the left panel) and $\tan \beta = 10$ (the right panel) for $\lambda = 1$ are shown:
\begin{equation}
\begin{split}
&\mathrm{(i)} \begin{cases}
\,m_1 = m_2 = 300 \, \mathrm{GeV} \\
\,M_1 = M_2 = 300 \, \mathrm{GeV} \\
\,m_{X_0} = m_{X_1} = m_{X_2} = 300 \, \mathrm{GeV},
\end{cases} \\
&\mathrm{(ii)} \begin{cases}
\,m_1 = m_2 = 500 \, \mathrm{GeV} \\
\,M_1 = M_2 = 300 \, \mathrm{GeV} \\
\,m_{X_0} = m_{X_1} = m_{X_2} = 300 \, \mathrm{GeV},
\end{cases} \\
&\mathrm{(iii)} \begin{cases}
\,m_1 = m_2 = 300 \, \mathrm{GeV} \\
\,M_1 = M_2 = 500 \, \mathrm{GeV} \\
\,m_{X_0} = m_{X_1} = m_{X_2} = 500 \, \mathrm{GeV},
\end{cases} \\
&\mathrm{(iv)} \begin{cases}
\,m_1 = m_2 = 300 \, \mathrm{GeV} \\
\,M_1 = M_2 = 300 \, \mathrm{GeV} \\
\,m_{X_0} = 0, \,\, m_{X_1} = m_{X_2} = 300 \, \mathrm{GeV}.
\end{cases} \\
\end{split}
\end{equation}
Here, $M_1$ and $M_2$ are the soft SUSY breaking masses of the Bino and
the Winos discussed in the next subsection (see \eqref{gmass}).
In the table, $m_{\sigma 1}, m_{\sigma 2}, m_{\sigma 3}$ denote (the square roots of) the three eigenvalues of the mass matrix \eqref{sigmass}, while $m_{\rho 1}, m_{\rho 2}, m_{\rho 3}$ are those of the imaginary parts $\rho$.
\begin{table}
\begin{minipage}{0.5\hsize}
\begin{center}
\begin{tabular}{|r||c|c|c|c|}
\hline
 & (i) & (ii) & (iii) & (iv) \\
\hline
$m_{\tilde{C}_1}$ [GeV] & 248 & 289 & 288 & 248 \\
\hline
$m_{\tilde{C}_2}$ [GeV] & 302 & 501 & 302 & 302 \\
\hline
$m_{\tilde{C}_3}$ [GeV] & 361 & 518 & 519 & 361 \\
\hline
$m_{\tilde{N}_1}$ [GeV] & 0 & 0 & 0 & 0 \\
\hline
$m_{\tilde{N}_2}$ [GeV] & 239 & 286 & 282 & 239 \\
\hline
$m_{\tilde{N}_3}$ [GeV] & 300 & 300 & 305 & 300 \\
\hline
$m_{\tilde{N}_4}$ [GeV] & 307 & 505 & 347 & 307 \\
\hline
$m_{\tilde{N}_5}$ [GeV] & 347 & 519 & 347 & 347 \\
\hline
$m_{\tilde{N}_6}$ [GeV] & 347 & 529 & 500 & 347 \\
\hline
$m_{\tilde{N}_7}$ [GeV] & 368 & 529 & 524 & 368 \\
\hline
$m_{X^-_1}$ [GeV] & 426 & 585 & 585 & 426 \\
\hline
$m_{X^+_2}$ [GeV] & 422 & 582 & 582 & 422 \\
\hline
$m_{\sigma 1}$, $m_{\rho 1}$ [GeV] & 301 & 300 & 500 & 119 \\
\hline
$m_{\sigma 2}$, $m_{\rho 2}$ [GeV] & 421 & 581 & 581 & 421 \\
\hline
$m_{\sigma 3}$, $m_{\rho 3}$ [GeV] & 461 & 611 & 610 & 446 \\
\hline
\end{tabular}
\end{center}
\end{minipage}
\begin{minipage}{0.5\hsize}
\begin{center}
\begin{tabular}{|r||c|c|c|c|}
\hline
 & (i) & (ii) & (iii) & (iv) \\
\hline
$m_{\tilde{C}_1}$ [GeV] & 249 & 289 & 289 & 249 \\
\hline
$m_{\tilde{C}_2}$ [GeV] & 300 & 500 & 300 & 300 \\
\hline
$m_{\tilde{C}_3}$ [GeV] & 362 & 519 & 519 & 362 \\
\hline
$m_{\tilde{N}_1}$ [GeV] & 0 & 0 & 0 & 0 \\
\hline
$m_{\tilde{N}_2}$ [GeV] & 239 & 286 & 282 & 239 \\
\hline
$m_{\tilde{N}_3}$ [GeV] & 300 & 300 & 305 & 300 \\
\hline
$m_{\tilde{N}_4}$ [GeV] & 307 & 505 & 347 & 307 \\
\hline
$m_{\tilde{N}_5}$ [GeV] & 347 & 519 & 347 & 347 \\
\hline
$m_{\tilde{N}_6}$ [GeV] & 347 & 529 & 500 & 347 \\
\hline
$m_{\tilde{N}_7}$ [GeV] & 368 & 529 & 524 & 368 \\
\hline
$m_{X^-_1}$ [GeV] & 427 & 585 & 585 & 427 \\
\hline
$m_{X^+_2}$ [GeV] & 422 & 581 & 581 & 422 \\
\hline
$m_{\sigma 1}$, $m_{\rho 1}$ [GeV] & 302 & 301 & 501 & 119 \\
\hline
$m_{\sigma 2}$, $m_{\rho 2}$ [GeV] & 420 & 580 & 580 & 420 \\
\hline
$m_{\sigma 3}$, $m_{\rho 3}$ [GeV] & 461 & 611 & 611 & 447 \\
\hline
\end{tabular}
\end{center}
\end{minipage}
\caption{The chargino masses, the neutralino masses, and the scalar masses of the $X$ fields in the four cases (i), (ii), (iii), (iv) explained in the main text. The left panel is plotted with $\tan \beta = 3$, while the right panel is done with $\tan \beta = 10$.}
\label{tab1}
\end{table}

\subsection{The fermion masses}

Let us analyze the fermion masses of the Higgs sector. We assume that
the gauginos have the following Majorana mass terms, which break
$U(1)_{R}$ symmetry softly
(presumably due to hidden sector dynamics):
\begin{equation}
\begin{split}
\mathcal{L}_{gaugino} &= - \frac{1}{2} M_2 \left( \tilde{W}^+ \tilde{W}^- + \tilde{W}^- \tilde{W}^+ \right) \\
&\quad - \frac{1}{2} M_2 \left( \tilde{W}^3 \tilde{W}^3 + c.c. \right) \\
&\quad - \frac{1}{2} M_1 \left( \tilde{B} \tilde{B} + c.c. \right),
\end{split}\label{gmass}
\end{equation}
where we have omitted the Gluino mass terms with no need for the present purposes.
Certain modes in the Higgs sector have the mass mixings with the Winos
and the Bino. These mixings come from the Yukawa-type coupling of a
gaugino, a fermion, and its scalar superpartner which has a nonzero
vacuum expectation value. The mixing terms of the Winos and the Higgsinos are given by
\begin{equation}
\begin{split}
\mathcal{L}_{SU(2)} &= - \sqrt{2} m_W \sin \beta \tilde{H}_u^+ \tilde{W}^- - \sqrt{2} m_W \cos \beta \tilde{H}_d^- \tilde{W}^+ \\
&\quad + m_Z \sin \beta \cos \theta_W \tilde{H}_u^0 \tilde{W}^3 - m_Z \cos \beta \cos \theta_W \tilde{H}_d^0 \tilde{W}^3 + c.c.,
\end{split}\label{2mix}
\end{equation}
while the mixing terms of the Bino and the Higgsinos are given by
\begin{equation}
\mathcal{L}_{U(1)} = - m_Z \sin \beta \sin \theta_W \tilde{H}_u^0 \tilde{B} + m_Z \cos \beta \sin \theta_W \tilde{H}_d^0 \tilde{B} + c.c.
\label{1mix}
\end{equation}

With the aid of these mass terms and the superpotential \eqref{visibleW}
of the model, we can derive the chargino mass terms, which are expressed as
\begin{equation}
\mathcal{L}_{chargino} = - \frac{1}{2} \psi^T \mathcal{M}_{\tilde{C}} \psi + c.c.,
\end{equation}
where $\psi = ( \tilde{W}^+, \tilde{H}_u^+, \tilde{X}_2^+, \tilde{W}^-, \tilde{H}_d^-, \tilde{X}_1^- )$ and the mass matrix $\mathcal{M}_{\tilde{C}}$ is given by
\begin{equation}
\mathcal{M}_{\tilde{C}} = \left(
  \begin{array}{cc}
  0 & M^T \\
  M & 0
  \end{array}
 \right), \quad
M = \left(
  \begin{array}{ccc}
  M_2 & \sqrt{2} m_W \sin \beta & 0 \\
  \sqrt{2} m_W \cos \beta &0 & m_2 \\
  0 & m_1 & 0
  \end{array}
 \right).
\end{equation}
The mass matrix $M$ can be diagonalized as
\begin{equation}
L^\ast M R^\dagger = \left(
  \begin{array}{ccc}
  m_{\tilde{C}_1} & 0 & 0 \\
  0 & m_{\tilde{C}_2} & 0 \\
  0 & 0 & m_{\tilde{C}_3}
  \end{array}
 \right),
\end{equation}
with the corresponding mass eigenstates given by
\begin{equation}
\left(
  \begin{array}{c}
  \tilde{C}_1^+ \\
  \tilde{C}_2^+ \\
  \tilde{C}_3^+
  \end{array}
 \right) = R \left(
  \begin{array}{c}
  \tilde{W}^+ \\
  \tilde{H}_u^+ \\
  \tilde{X}_2^+
  \end{array}
 \right), \quad
\left(
  \begin{array}{c}
  \tilde{C}_1^- \\
  \tilde{C}_2^- \\
  \tilde{C}_3^-
  \end{array}
 \right) = L \left(
  \begin{array}{c}
  \tilde{W}^- \\
  \tilde{H}_d^- \\
  \tilde{X}_1^-
  \end{array}
 \right),
\end{equation}
where $L$ and $R$ are unitary matrices. We present sample mass spectra
of the charginos in table~\ref{tab1}, where we show four cases (i),
(ii), (iii), (iv) given above with $\tan \beta = 3$ (the left panel) and
$\tan \beta = 10$ (the right panel).

Finally, we analyze the masses of the neutralinos $\psi^0 = ( \tilde{B}, \tilde{W}^3, \tilde{H}_d^0, \tilde{H}_u^0, \tilde{X}_1^0, \tilde{X}_2^0, \tilde{X}_0 )$. The neutralino mass terms are expressed as
\begin{equation}
\mathcal{L}_{neutralino} = - \frac{1}{2} (\psi^0)^T \mathcal{M}_{\tilde{N}} \psi^0 + c.c.,
\end{equation}
where the mass matrix $\mathcal{M}_{\tilde{N}}$ is given by
\begin{equation}
\mathcal{M}_{\tilde{N}} =
\left(
  \begin{array}{ccccccc}
  M_1 & 0 & - m_Z c_{\beta} s_W & m_Z s_{\beta} s_W & 0 & 0 & 0 \\
  0 & M_2 & m_Z c_\beta c_W & - m_Z s_\beta c_W & 0 & 0 & 0 \\
  - m_Z c_\beta s_W &  m_Z c_\beta c_W & 0 & 0 & 0 & - m_2 & -\frac{\lambda}{\sqrt{2}} v s_\beta \\
   m_Z s_\beta s_W &  - m_Z s_\beta c_W & 0 & 0 & - m_1 & 0 & -\frac{\lambda}{\sqrt{2}} v c_\beta \\
   0 & 0 & 0 & - m_1 & 0 & 0 & 0 \\
   0 & 0 & - m_2 & 0 & 0 & 0 & 0 \\
   0 & 0 & -\frac{\lambda}{\sqrt{2}} v s_\beta & -\frac{\lambda}{\sqrt{2}} v c_\beta & 0 & 0 & 0 \\
  \end{array}
 \right),
\label{nmass}
\end{equation}
with $s_W=\sin \theta_W$ and $c_W=\cos \theta_W$.
We also present sample mass spectra of the neutralinos in table~\ref{tab1},
where we show four cases (i), (ii), (iii), (iv) given above with $\tan \beta
= 3$ (the left panel) and $\tan \beta = 10$ (the right panel) for $\lambda = 1$.
The table implies that the lightest neutralino $\tilde{N}_1$ is massless at
the tree-level. This originates from the fact that the determinant of the mass matrix
\eqref{nmass} is vanishing:
\begin{equation}
\det \mathcal{M}_{\tilde{N}} = 0.
\end{equation}
The massless mode would correspond to the goldstino mode in the visible SUSY breaking
without soft SUSY breaking terms.
In the full setup with hidden sector SUSY breaking in supergravity, it
turns out to be a massive pseudo-goldstino.
If the hidden sector was sequestered from our visible sector,\footnote{In this case, sleptons have tachyonic 
masses due to the anomaly mediation \cite{Randall:1998uk}.}
the pseudo-goldstino mass would be twice the gravitino mass in accord with 
\cite{Cheung:2010mc}, whereas it may be orders of magnitude different
from the gravitino mass in general due to higher order effects beyond the
simple tree-level analysis.
Anyhow, the visible sector pseudo-goldstino might be seen as a
remarkable feature in the visible SUSY breaking scenario.%
\footnote{Such a low-scale pseudo-goldstino, as well as gravitinos (see also footnote
\ref{fn:extended}), might constitute extra radiation
\cite{Nak} in the early universe.}
Of course, we have more to investigate on higher order effects.
For instance, the visible SUSY breaking in the Higgs sector also affects soft SUSY
breaking pattern due to gauge mediation effects with the Higgses as messengers
\cite{Evans:2010kd}. These and other features depend crucially
on the hidden sector SUSY breaking and its mediation to the visible
sector, in particular, its connection with Higgs interactions thereof.

\section{A connection with hidden sector SUSY breaking}

As an illustrative example of connecting the visible SUSY breaking in the
Higgs sector to the higher-scale SUSY breaking in the hidden sector,
we here present Giudice-Masiero-like effective operators
\cite{Giudice:1988yz} in our setup.
This also serves as a sample case that the
visible SUSY breaking is a cascade phenomenon induced by the hidden
sector SUSY breaking.
Let us consider both $F$-type and $D$-type SUSY
breaking spurions representing the hidden sector effects:
\begin{equation}
S = \theta^2 F, \qquad W_{\alpha} = \theta_{\alpha} D,
\end{equation}
whose R-charges are $2$ and $1$, respectively.

Then the superpotential \eqref{visibleW} comes from
\begin{equation}
\begin{split}
\int d^4 \theta \,\, \biggl[ a_1 \frac{S^\dagger}{M} X_1 H_u + a_2 \frac{S^\dagger}{M} X_2 H_d
+ a_0 \frac{\overline{W}_{\dot{\alpha}} \overline{W}^{\dot{\alpha}}}{M^2} X_0 + c.c. \biggr],
\end{split}
\end{equation}
where $M$ is the mediation scale of the hidden sector SUSY breaking to
the Higgs sector with $a_0, a_1, a_2$ as coupling constants.%
\footnote{We have not included a term like $S^\dagger X_0$ without $M$
suppression. If the $S$ has a non-vanishing scalar component, we may do
without the $D$-type SUSY breaking spurion by 
replacing the $W_\alpha$-dependent term with a term like $S^{\dagger 2}SX_0/M^2$.}
These terms result in the parameters
\begin{equation}
m_1 = a_1 \frac{F^\dagger}{M}, \quad m_2 = a_2 \frac{F^\dagger}{M}, \quad f = a_0 \frac{D^2}{M^2}.
\end{equation}
We can also obtain the soft SUSY breaking terms with the aid of terms like
\begin{equation}
\begin{split}
&\int d^4 \theta \,\, \biggl[ \left( \frac{S^\dagger S}{M^2} H_u H_d + c.c. \right) \\
&+ \frac{S^\dagger S}{M^2} \left( H_u^\dagger H_u + H_d^\dagger H_d + X_0^\dagger X_0 + X_1^\dagger X_1 + X_2^\dagger X_2 \right) \biggr],
\end{split}
\end{equation}
where the first term gives the $B\mu$-term
and the rest gives soft scalar masses of the Higgs sector in our model.

\section{Conclusion}

We have presented a supersymmetric extension of the
standard model whose Higgs sector has spontaneous SUSY breaking even in
the absence of the soft breaking terms from the usual hidden
sector. This extension is along the lines of general perspectives such
that the Higgs sector may be a window to some unknown physics
and SUSY breaking may be ubiquitous even in the visible sector.
The current experimental 
bound for the lighter CP-even Higgs mass can be evaded even at the
tree-level, which is reminiscent of the NMSSM. The pseudo-goldstino lies in the visible sector
since the corresponding SUSY breaking is visible in the Higgs sector.

Since the scale of the visible SUSY breaking can be near the EW scale, it
might be possible to observe the breaking dynamics 
rather directly in future experiments. 
It may be interesting to analyze new decay channels of the Higgs fields
in such a model. We have regarded the current experimental bound $m_h > 114$ GeV 
for the lighter CP-even Higgs field as a point of reference in our consideration. 
However, this bound might be totally inadequate for our model since, among others,
decays of Higgs particles beyond the standard model have not been taken into account.
In this connection, the
production and detection of the pseudo-goldstino mode
in the Higgs sector is another interesting experimental challenge.

We have restricted ourselves to the vacuum in our model
that has a desired breaking pattern of the visible SUSY and the EW
symmetry in this paper.
In the MSSM and its cousins, 
thorough analyses of the potentially dangerous
directions in their field spaces have been carried out
\cite{Frere:1983ag}. Our extension might have
charge and/or color breaking minima in the landscape of vacua,
which is to be further examined. 

We have not specified the details of the hidden sector SUSY breaking
in the present analyses mainly at the tree-level,
though it is intriguing to study connections between the visible SUSY
breaking in the Higgs sector and the hidden sector SUSY breaking in a
variety of mediation mechanisms.
Since the Higgs sector is largely unknown experimentally,
and even theoretically, we often encounter puzzles such as $\mu$ and $B\mu$ problems
in the MSSM with hidden sector SUSY breaking, various possibilities concerning the Higgs sector and its
possible extensions may deserve open-minded investigations.

\section*{Acknowledgments}

We would like to thank M.~Sakai for discussions.
I.~K.-I. would like to acknowledge discussions with F.~Takahashi
and T.T.~Yanagida.
T.~S. is the Yukawa Fellow and the work of T.~S. is partially supported by Yukawa Memorial Foundation.
This work is supported by the Grant-in-Aid for Yukawa International
Program for Quark-Hadron Sciences, the Grant-in-Aid
for the Global COE Program "The Next Generation of Physics,
Spun from Universality and Emergence", and
World Premier International Research Center Initiative
(WPI Initiative), MEXT, Japan.

\section*{Appendix}

Here, we summarize the mass terms of the Higgs sector fields derived from the scalar potential \eqref{visibleV}. We can read the mass matrices presented in the main text from these mass terms.

\subsection*{The charged Higgs mass terms}

The mass terms from the $F$-term contribution to the scalar potential
$V_F$ are given by
\begin{equation*}
m_1^2 |H_u^+|^2 + m_2^2 |H_d^-|^2 +\lambda \left( f- \frac{1}{2} \lambda v^2 \sin \beta \cos \beta \right) \left( H_u^+ H_d^- + c.c. \right),
\end{equation*}
where we have used the vacuum expectation values of the Higgs fields
$H_u^0 = \frac{1}{\sqrt{2}} v \sin \beta$, $H_d^0 = \frac{1}{\sqrt{2}} v
\cos \beta$, and the redefinitions of the scalar fields \eqref{shift}.
The $D$-term contribution to the charged Higgs mass terms is given by
\begin{equation*}
\begin{split}
&-\frac{1}{8} g^2 v^2 \cos 2\beta \left( | H_u^+ |^2 - | H_d^- |^2 \right) +\frac{1}{4} g_2^2 v^2 \cos^2 \beta | H_u^+ |^2 +\frac{1}{4} g_2^2 v^2 \sin^2 \beta | H_d^- |^2 \\
&\quad +\frac{1}{4} g_2^2 v^2 \sin \beta \cos \beta \left( H_u^+ H_d^- + c.c. \right).
\end{split}\label{chargedDmass}
\end{equation*}
The contribution from the soft SUSY breaking terms is given by
\begin{equation*}
m_{H_u}^2 | H_u^+ |^2 + m_{H_d}^2 | H_d^- |^2 + b \left( H_u^+ H_d^- + c.c. \right).
\end{equation*}

\subsection*{The neutral Higgs mass terms}

The mass terms from the $F$-term contribution to the scalar potential
$V_F$ are given by
\begin{equation*}
\begin{split}
&\left(m_1^2 +  \frac{1}{2} \lambda^2 v^2 \cos^2 \beta \right) |H_u^0|^2 + \left(m_2^2 +  \frac{1}{2} \lambda^2 v^2 \sin^2 \beta \right) |H_d^0|^2 \\
&+ \frac{1}{2} \lambda^2 v^2 \sin \beta \cos \beta \left( H_u^0 {H_d^0}^\ast + c.c. \right) \\
&-\lambda \left( f- \frac{1}{2} \lambda v^2 \sin \beta \cos \beta \right) \left( H_u^0 H_d^0 + c.c. \right).
\end{split}
\end{equation*}
The contribution from the $D$-term $V_{D}$ is given by
\begin{equation*}
\begin{split}
&\frac{1}{8} g^2 \biggl[ v^2 \left( \sin^2 \beta - \cos^2 \beta \right) \left( | H_u^0 |^2 - | H_d^0 |^2 \right) \\
&\quad +\frac{1}{2} \left( \sin \beta \left(H_u^0 + {H_u^0}^\ast \right) - \cos \beta \left(H_u^0 + {H_u^0}^\ast \right) \right)^2 \biggr].
\end{split}
\end{equation*}
The soft SUSY breaking contribution is given by
\begin{equation*}
m_{H_u}^2 | H_u^0 |^2 + m_{H_d}^2 | H_d^0 |^2 - b \left( H_u^0 H_d^0 + c.c. \right).
\end{equation*}

\subsection*{The scalar mass terms of the $X$ fields}

The mass terms of the $X$ scalar fields are given by

\begin{equation*}
\begin{split}
&\left( \frac{1}{2} \lambda^2 v^2 + m_{X_0}^2 \right) |X_0|^2 \\
&+\left( m_1^2 + m_{X_1}^2 + \frac{1}{2} m_Z^2 \cos 2\beta \right) |X_1^0|^2
+\left( m_2^2 + m_{X_2}^2 - \frac{1}{2} m_Z^2 \cos 2\beta \right) |X_2^0|^2 \\
&+\left( m_1^2 + m_{X_1}^2 - \frac{1}{2} m_Z^2 \cos 2\theta_W \cos 2\beta \right) |X_1^-|^2 \\
&+\left( m_2^2 + m_{X_2}^2 + \frac{1}{2} m_Z^2 \cos 2\theta_W \cos 2\beta \right) |X_2^+|^2 \\
&-\frac{\lambda}{\sqrt{2}} m_1 v \cos \beta \left( X_0^\ast X_1^0 +  X_0 {X_1^0}^\ast \right)
+\frac{\lambda}{\sqrt{2}} m_2 v \sin \beta \left( X_0^\ast X_2^0 +  X_0 {X_2^0}^\ast \right).
\end{split}
\end{equation*}

%
%


\begin{thebibliography}{1}

\bibitem{Martin:1997ns}
  For a review, S.~P.~Martin,
  arXiv:hep-ph/9709356.

\bibitem{Ibe}
  See M.~Ibe, A.~Rajaraman and Z.~Surujon,
  arXiv:1012.5099 [hep-ph].

\bibitem{Nel}
  See A.~E.~Nelson, N.~Rius, V.~Sanz and M.~Unsal,
  JHEP {\bf 0208}, 039 (2002) [arXiv:hep-ph/0206102].

\bibitem{NMSSM}
  For recent reviews,
  M.~Maniatis,
  Int.\ J.\ Mod.\ Phys.\  A {\bf 25}, 3505 (2010)
  [arXiv:0906.0777 [hep-ph]];
  U.~Ellwanger, C.~Hugonie and A.~M.~Teixeira,
  Phys.\ Rept.\  {\bf 496}, 1 (2010)
  [arXiv:0910.1785 [hep-ph]].

\bibitem{deGouvea:1997cx}
  See A.~de Gouvea, A.~Friedland and H.~Murayama,
  Phys.\ Rev.\  D {\bf 57}, 5676 (1998)
  [arXiv:hep-ph/9711264].

\bibitem{Okada:1990gg}
  Y.~Okada, M.~Yamaguchi and T.~Yanagida,
  Phys.\ Lett.\  B {\bf 262}, 54 (1991);
  H.~E.~Haber and R.~Hempfling,
  Phys.\ Rev.\ Lett.\  {\bf 66}, 1815 (1991);
  J.~R.~Ellis, G.~Ridolfi and F.~Zwirner,
  Phys.\ Lett.\  B {\bf 262}, 477 (1991).

\bibitem{Agashe:1997kn}
  K.~Agashe and M.~Graesser,
  Nucl.\ Phys.\  B {\bf 507}, 3 (1997)
  [arXiv:hep-ph/9704206];
  A.~Delgado, G.~F.~Giudice and P.~Slavich,
  Phys.\ Lett.\  B {\bf 653}, 424 (2007)
  [arXiv:0706.3873 [hep-ph]];
  G.~F.~Giudice, H.~D.~Kim and R.~Rattazzi,
  Phys.\ Lett.\  B {\bf 660}, 545 (2008)
  [arXiv:0711.4448 [hep-ph]];
  T.~Liu and C.~E.~M.~Wagner,
  JHEP {\bf 0806}, 073 (2008)
  [arXiv:0803.2895 [hep-ph]];
  U.~Ellwanger, C.~C.~Jean-Louis and A.~M.~Teixeira,
  JHEP {\bf 0805}, 044 (2008)
  [arXiv:0803.2962 [hep-ph]].

\bibitem{lh}
  S.~Abel, M.~J.~Dolan, J.~Jaeckel and V.~V.~Khoze,
  JHEP {\bf 0912}, 001 (2009)
  [arXiv:0910.2674 [hep-ph]];
  T.~Kobayashi, Y.~Nakai and R.~Takahashi,
  JHEP {\bf 1001}, 003 (2010)
  [arXiv:0910.3477 [hep-ph]].

\bibitem{Meade:2008wd}
  P.~Meade, N.~Seiberg and D.~Shih,
  Prog.\ Theor.\ Phys.\ Suppl.\  {\bf 177}, 143 (2009)
  [arXiv:0801.3278 [hep-ph]].

\bibitem{Patt}
  See B.~Patt and F.~Wilczek,
  arXiv:hep-ph/0605188.

\bibitem{Kumar}
  See P.~Kumar and J.~D.~Lykken,
  JHEP {\bf 0407}, 001 (2004) [arXiv:hep-ph/0401140].

\bibitem{INY}
  Izawa K.-I., Y.~Nomura and T.~Yanagida,
  Prog.\ Theor.\ Phys.\ {\bf 102} 1181 (1999)
  [arXiv:hep-ph/9908240].

\bibitem{Izawa:2010bc}
  Izawa K.-I., Y.~Nakai and R.~Takahashi,
  Phys.\ Rev.\  D {\bf 82}, 075008 (2010)
  [arXiv:1003.0740 [hep-ph]].

\bibitem{Kribs:2007ac}
  G.~D.~Kribs, E.~Poppitz and N.~Weiner,
  Phys.\ Rev.\  D {\bf 78}, 055010 (2008)
  [arXiv:0712.2039 [hep-ph]].

\bibitem{Amaldi:1991cn}
  U.~Amaldi, W.~de Boer and H.~Furstenau,
  Phys.\ Lett.\  B {\bf 260}, 447 (1991).

\bibitem{gauge}
  See M.~Dine and J.~Mason,
  Phys.\ Rev.\  D {\bf 77}, 016005 (2008)
  [arXiv:hep-ph/0611312];
  Y.~Nakai and Y.~Ookouchi,
  arXiv:1010.5540 [hep-th].

\bibitem{Batra:2008rc}
  See P.~Batra and E.~Ponton,
  Phys.\ Rev.\  {\bf D79}, 035001 (2009)
  [arXiv:0809.3453 [hep-ph]].

\bibitem{Randall:1998uk}
  L.~Randall and R.~Sundrum,
  Nucl.\ Phys.\  {\bf B557}, 79-118 (1999).
  [hep-th/9810155];
  G.~F.~Giudice, M.~A.~Luty and H.~Murayama {\it et al.},
  JHEP {\bf 9812}, 027 (1998).
  [hep-ph/9810442].

\bibitem{Cheung:2010mc}
  C.~Cheung, Y.~Nomura and J.~Thaler,
  JHEP {\bf 1003}, 073 (2010)
  [arXiv:1002.1967 [hep-ph]].
%
  See also N.~Craig, J.~March-Russell and M.~McCullough,
  JHEP {\bf 1010}, 095 (2010)
  [arXiv:1007.1239 [hep-ph]].

\bibitem{Nak}
  See K.~Nakayama, F.~Takahashi and T.~T.~Yanagida,
  arXiv:1010.5693 [hep-ph], and references therein.

\bibitem{Evans:2010kd}
  See J.~L.~Evans, M.~Sudano and T.~T.~Yanagida,
  arXiv:1012.2952 [hep-ph].

\bibitem{Giudice:1988yz}
  G.~F.~Giudice and A.~Masiero,
  Phys.\ Lett.\  B {\bf 206}, 480 (1988).
  See also M.~McCullough,
  Phys.\ Rev.\  D {\bf 82}, 115016 (2010)
  [arXiv:1010.3203 [hep-ph]].

\bibitem{Frere:1983ag}
  J.~M.~Frere, D.~R.~T.~Jones and S.~Raby,
  Nucl.\ Phys.\  B {\bf 222}, 11 (1983);
  L.~Alvarez-Gaume, J.~Polchinski and M.~B.~Wise,
  Nucl.\ Phys.\  B {\bf 221}, 495 (1983);
  J.~P.~Derendinger and C.~A.~Savoy,
  Nucl.\ Phys.\  B {\bf 237}, 307 (1984);
  J.~A.~Casas, A.~Lleyda and C.~Munoz,
  Nucl.\ Phys.\  B {\bf 471}, 3 (1996)
  [arXiv:hep-ph/9507294], and references therein;
  T.~Kobayashi and T.~Shimomura,
  Phys.\ Rev.\  D {\bf 82}, 035008 (2010)
  [arXiv:1006.0062 [hep-ph]];
  Y.~Kanehata, T.~Kobayashi, Y.~Konishi and T.~Shimomura,
  Phys.\ Rev.\  D {\bf 82}, 075018 (2010)
  [arXiv:1008.0593 [hep-ph]].

\end{thebibliography}
\end{document}